\documentclass[onecollarge,draft]{svjour2}

\usepackage{bm}

\begin{document}


\title{An Approach to Loop Quantum Cosmology Through
Integrable Discrete Heisenberg Spin Chains}

\author{Christine C. Dantas}

 \institute{Divis\~ao de Materiais (AMR), Instituto de Aeron\'autica e Espa\c co (IAE), Departamento de Ci\^encia e Tecnologia Aeroespacial (CTA), Brazil \email{christineccd@iae.cta.br}}

\journalname{Foundations of Physics}

\date{Submitted: \today}

\maketitle

\begin{abstract}
The quantum evolution equation of Loop Quantum Cosmology (LQC) -- the quantum Hamiltonian constraint -- is a difference equation. We relate the LQC constraint equation in vacuum Bianchi I separable (locally rotationally symmetric) models with an integrable differential-difference nonlinear Schr\"odinger type equation, which in turn is known to be  associated with integrable, discrete Heisenberg spin chain models in condensed matter physics. We illustrate the similarity between both systems with a simple constraint in the linear regime.
\end{abstract}



\large

\section{Introduction}

Loop Quantum Gravity (LQG) \cite{LQG} aims to describe the physics of spacetime at Planck-length scales. In the last few years, research in this area has progressed towards predictions that area and volume observables are fundamentally quantized. In addition, LQG has been explored in symmetry reduced models, which take in consideration simple cosmological spacetimes, a research area known as Loop Quantum Cosmology (LQC) \cite{LQC}. 

The quantum evolution equation of LQG -- the quantum Hamiltonian constraint -- is a difference equation for the wavefunction (i.e., a discrete recursion relation), instead of a differential equation like the Wheeler-de Witt equation (see, e.g., Ref. \cite{Ash09}). A series of consistency studies indicates that solutions to the quantum evolution equation, under certain conditions, do not lead to singularities: a minimum observable volume leads to a ``bounce" near classical singularities and quantum states evolve deterministically through those singularities (e.g., \cite{Ash08}, \cite{Ash09}). Recently, the nature of the bounce and its robustness have been addressed under several approaches: numerically; by use of approximation methods (``effective equation" techniques); or by the analysis of simplified exact analytical solutions (for reviews, see, e.g., \cite{Sin12}, \cite{Ash11} and \cite{Boj12} and references therein).

However, as symmetry-reduced versions of LQG, LQC models assume certain results directly from full LQG. For instance, the smallest nonzero area eigenvalue of LQG is the assumed step size (the so called ``area gap") of the LQC difference equation. It has been shown that the Wheeler-de Witt equation can be made to agree with LQC (in the case of a $k=0$, $\Lambda=0$ FRW cosmological model, coupled to a massless scalar field), as the area gap diminishes. However, the approximation is not uniform in the chosen interval of ``internal time", and in fact if the area gap is set to zero, LQC does not admit any limit, being an intrinsically discrete theory \cite{Ash08}. An elucidation on the nature of the solutions of the discrete Hamiltonian constraint in more general cases is necessary.

Here, we propose a technique which may be able to address these questions more generally, by observing that solutions to the fundamental equation of micromagnetism in condensed matter have been explored in the context of similar difference equations. 

\section{LQC and Nonlinear Schr\"odinger equations}

Generally, the understanding of the magnetization vector field dynamics of a submicron magnetic particle is made at at the continuum approximation, according to the Landau-Lifshitz equation (LL; also known as the Landau-Lifshitz-Gilbert equation with the inclusion of damping) \cite{LL}. It describes the local precessional motion of the magnetization vector under an effective magnetic field which represents several distinct interactions amongst the spins. Being a highly nonlinear vector partial differential equation, the LL equation is generally solved by using numerical techniques (analytical solutions can only be obtained in very few  cases; see, e.g., \cite{LLAS}). In particular, the dynamics of classical Heisenberg ferromagnetic spin chain models is governed by the LL equation. Such models are interesting nonlinear systems, displaying a rich diversity of dynamical properties, including soliton spin excitations \cite{SCM}. 

It has been realized, however, an interesting relation between such models and a family of completely integrable nonlinear Schr\"odinger (NLS) equations \cite{Lak}. For instance, it can be shown that, given the one-dimensional isotropic ferromagnetic spin chain classical Hamiltonian $H = - \sum_n \vec{S}_n \cdot \vec{S}_{n+1}$ (wherein the resulting LL equation of motion is $\partial \vec{S}/\partial t = \vec{S} \times \partial^2\vec{S}/\partial x^2$), there is an associated cubic NLS equation, which is completely integrable, given by $i(\partial q /\partial t) + (\partial^2 q/\partial x^2) + 2 |q|^2 q = 0$. These results were obtained by the use of integrability methods such as the gauge equivalence and the geometric equivalence (space curve) formalisms (\cite{Dol94} and references therein). However, a more realistic description of ferromagnetic systems is a formulation of the dynamics at the lattice level, instead of the continuum limit. The integrability of such discrete systems has also been addressed under the methods above, suitably adapted for discretization \cite{Dol95}, \cite{Dan98}.

In the present paper, we present a simple example which makes use of the above mentioned results of discrete Heisenberg ferromagnetic spin chain models in the context of difference equations of LQC models. We will focus on a simple LQC evolution equation in vacuum Bianchi I separable, LRS (locally rotationally symmetric) models, which are the simplest anisotropic model (with only two degrees of freedom), particularly we shall focus only on the resulting one-parameter, discrete recursion relation that arises from this model \cite{HC}:

\begin{equation}
\lambda f(m) q_{\lambda}(m) = q_{\lambda}(m+1) - q_{\lambda}(m-1),
\end{equation}
where $q(m)$ ($m$, real numbers) is related to the coefficients of the wavefunction $t_{m,n}$, which in turn are rescaled versions of the original wavefunction, normalized by a factor of the world volume of each basis state \cite{HC}. $\lambda$ is a separation parameter that arises in the scheme, and the function $f(m)$ ($f(m)\equiv 1/m$ for $m \geq 1$ and $f(m) = 0$ form $m = 0$) is a simplification of the function $d(m)$ in the last Ref. of \cite{HC}.

In parallel, we introduce the following completely integrable differential-difference nonlinear Schr\"odinger (NLS) equation (see, e.g., \cite{Dan98}):

\begin{equation}
i {dq(n) \over dt} = \left ( 1 + |q(n)|^2 \right )
\left [ q(n+1)+ q(n-1) \right ] - 2 q(n),
\end{equation}
where $n$ represents an integer sequence of points. We parametrize the factor $(1 + |q(n)|^2)$ in that equation by the operator $\hat{\alpha}$, and write the NLS equation as:
\begin{equation}
\hat{\mathcal{G}} q(n)= q(n+1)+ q(n-1), \label{G}
\end{equation}
where the operator $\hat{\mathcal{G}}$ is given by:
\begin{equation}
\hat{\mathcal{G}} \equiv {1 \over \hat{\alpha}} \left [ i {d \over dt} + 2 \right ]
\end{equation}
 We also rewrite the LQC equation, Eq. (1), as:
\begin{equation}
\hat{\mathcal{F}} q_{\lambda}(m) = q_{\lambda}(m+1) - q_{\lambda}(m-1), \label{F}
\end{equation}
with $\hat{\mathcal{F}} \equiv \lambda f(m)$. Whether the LQC and NLS discrete equations (Eqs. \ref{G} and \ref{F}) can be made mathematically equivalent in form is a relatively straightforward question, as far as one assumes some restrictions, whose validity on physical grounds is, nevertheless, unclear at this point. We shall address possible implications as we proceed. Eqs. \ref{G} and \ref{F} formally differ from each other in the following points: (i) the nature of operators $\hat{\mathcal{G}}$ and $\hat{\mathcal{F}}$, (ii) the nature of indexes $m$ and $n$ (iii) the existence of a family of equivalent equations parametrized further by the $\lambda$ in the LQC equation, and  (iv) the existence of a plus sign instead of minus sign in the right hand sides of these equations. Let us analyse each of these items separately.

\section{Linear sector}

First, equivalence between operators $\hat{\mathcal{G}}$ and $\hat{\mathcal{F}}$ impose the restriction that the wavefunctions $q_{\lambda}(m)$ should be of the form
\begin{equation}
q_{\lambda}(m) = e^{i [k_{\lambda}X(m)- \Phi_0]}, \label{Res1}
\end{equation}
where $k_{\lambda} \sim \hat{\alpha} \lambda $ defines a fiducial wavenumber; $X(m) \sim f(m)$, a fiducial sequence ``length", and $\Phi_0 \equiv 2$, a fixed phase. We notice here a similarity between that restriction and the choice of basis elements in Ref. \cite{Con06},
\begin{equation}
b_k(m) = e^{ik \exp{(-{2m \over M})}}, \label{basis}
\end{equation}
where $M$ a fiducial length. Such basis is adopted in that work in order to find pre-classical solutions to the separable LQC equation (Eq. \ref{F}). Pre-classicality is a concept introduced in order to discard unphysical solutions to the Hamiltonian constraint equations. These are solutions with highly oscillatory behavior arising far away from the corresponding classical singularity; in other words, pre-classical solutions are those which the coefficients of the wave function at large volume should not oscillate when small changes to the volume (e.g., by amounts of Planck size) are made (for a critical review on pre-classicality, see Section 5.18 of the first Ref. of \cite{LQC}). The idea in Ref. \cite{Con06} is to write a generic solution to Eq. \ref{F} as an expansion of the basis (\ref{basis}), which enables oscillatory behavior for small values of the index parameters, but at the same time singles out only smoother behavior for larger values of the parameters. We note that a choice of $X(m)$ in Eq. \ref{Res1} can be made to agree with the exponential factor in Eq. \ref{basis}, generally, as:
\begin{equation}
X(m) \sim f(m) \sim {1 \over m} \Leftrightarrow \exp{(-{2m \over M})} \simeq 1 - {2m \over M},
\end{equation}
which implies that $M$ is now a function of $m$,
\begin{equation}
M(m) \simeq - {2m^2 \over 1-m}, ~~~~~~~~ m > 1,
\end{equation}
in order that our restricted wavefunctions (\ref{Res1}) allow pre-classical solutions in the form considered in Ref. \cite{Con06}. It means that each $m$ in the sequence is associated with a given fiducial $M$. Other arbitrary forms can evidently be assumed for $X(m)$, which are completely independent on our assumptions concerning the equivalence between the LQC and NLS discrete equations (Eqs. \ref{G} and \ref{F}), since for the moment such equivalence in terms of the operators  $\hat{\mathcal{G}}$ and $\hat{\mathcal{F}}$ only restricts the form of the wavefunctions $q_{\lambda}(m)$ in accordance with Eq. \ref{Res1}.

Second, concerning the nature of the indexes, we only point out that pre-classical solutions to Eq. \ref{F} are known for both when $m$ is an integer or not, through the method of generating functions (c.f. \cite{HC}). For definiteness, we analyse only integer index parameters, in accordance with the integer index parameters of the NLS equation formulation (Eq. \ref{G}).
Third, concerning the parametrization of solutions according to $\lambda$, we assume they enter our formulation through the restriction \ref{Res1}, which can be freely varied for each family of solutions dictated by $\lambda$. Hence, we will assume that $\lambda$ is fixed {\it a priori} in order to proceed with the equivalence between Eqs. \ref{G} and \ref{F}.

Finally, we address freedom of signs (plus, minus) involved in the right hand sides of Eqs. \ref{G} and \ref{F}. For that, we briefly outline the method of space curve formalism (mentioned in the previous section), which allows one to map continuum Heisenberg spin-chain models onto the NLS family of equations. Our analysis was based on the results obtained by the discrete mapping procedure of Daniel and Manivannan \cite{Dan98}, motivated by the extension of the space curve formalism for discrete cases by Doliwa and Santini \cite{Dol95}. The reader is referred to these papers (and references therein) for details. 

The idea arises from well-known developments in mathematical physics, connecting the differential geometry of submanifolds and nonlinear partial differential equations. In essence, one looks for a fundamental geometric characterization of the motions of a curve in a submanifold which selects, among
all possible dynamics, integrable dynamics \cite{Dol94}. The particular method for the problem at hand begins by considering a discrete curve on a sphere, represented by a sequence of points and a set of three orthonormal basis vectors associated with each point (defined by a position vector with respect to a reference point at the center of the sphere). One also specifies adequate transition matrices from on basis to the next basis, in this case, these matrices can be combined into a rotation matrix depending only on two angles ($\theta_n, \phi_n$), namely, involving two of the basis vectors at subsequent points of the curve. The evolution of the basis is described by a matrix equation, and by defining a shift operator along the curve, as well as a compatibility condition for this operator, one reaches at a set of coupled evolution equation for the angles $\theta_n, \phi_n$. One then maps the Hamiltonian of the discrete chain model to be studied and its discrete equation of motion onto the discrete curve by a direct identification of the spin vector to one of the unit basis vectors (the radial vector). Such association leads to a completely equivalent set of coupled differential-difference equations for the evolution of the two angles $\theta_n, \phi_n$. The question at this point is to verify whether such coupled equations are integrable by finding a suitable transformation to a known integrable differential-difference equation.

By the method outline above, Daniel and Manivannan show that the completely integrable differential-difference NLS equation (Eq. \ref{G}) is associated with a classical equation of motion for the discrete Heisenberg spin chain, the Ishimori's model (based on Ishimori's partial differential equation, c.f. Ref. \cite{Ish82}),  which is a classical discrete equation of motion, given by:

\begin{equation}
{d \over dt} \vec{S}_n(t) =  2 \vec{S}_n \times
\left [
{\vec{S}_{n+1}\over 1 + \vec{S}_n \cdot \vec{S}_{n+1} } 
+ 
{\vec{S}_{n-1}\over 1 + \vec{S}_n \cdot \vec{S}_{n-1}}
\right ],
\end{equation}
where $\vec{S}_n = (S_x,S_y,S_z)$ are classical 3D spin vectors, with the constraint that $|\vec{S}_n|^2 = $ should be kept as a constant. 

Now returning to the question of the freedom of signs involved in the right hand sides of Eqs. \ref{G} and \ref{F}, by a detailed inspection of the equations of the mapping procedure outlined above, the LQC difference equation can also be mapped to Ishimori's model, inasmuch the coordinates are simply counter-rotated by a substitition $\sin \phi_n \rightarrow - \sin \phi_n$ in the matrix given by Eq. (6b) in that paper, in order to account now for the minus sign in the difference equation (\ref{F}). {\sl Therefore, the LQC difference equation presents a completely equivalent discrete spin chain equation counterpart, mathematically equivalent to the Ishimori's model, and therefore, under the conditions previously stated, it is integrable.}

\section{Conclusions}

It is interesting to note that the equivalent counterpart of the complete Ishimori's equation \cite{Ish82} (spin-one field model {\sl in the plane}) is given formally the Davey-Stewartson equation \cite{Dav74}, which describes modulated nonlinear surface gravity waves, propagating over a horizontal sea bed (3D packets of mechanical surface waves)\footnote{In 1+1 dimensions, this equation reduces to a simple nonlinear Schr\"odinger equation, Eq. \ref{G} of the present paper.}. We speculate to what extent these models are somehow able to mimmic other classes of reduced LQC models, and if so, what they are telling us about the propagation through classical singularities.

In summary, we have raised the possibility of relating the LQC constraint equation in vacuum Bianchi I separable models with an integrable differential-difference nonlinear Schr\"odinger type equation, which in turn is known to be  associated with integrable, discrete Heisenberg spin chain models. If this association proves to be correct at some level, then it is encouraging to have integrable equations at our disposal as LQC toy models, specially in the nonlinear regime. Here we have only pointed out an illustration in the linear regime. This proposal also opens the possibility of studying the propagation of quantum gravity reduced wavefunctions models using spin chain representation models, which are amenable for computer simulations. Several already available numerical integrators of micromagnetism can be adapted to study these models. However, a deeper understanding of the physical meaning of ``Ishimori-like quantum gravity models", specially the meaning of the vector spin, are still unclear at this point, and research in this area is under way.

\begin{acknowledgement}

The author acknowledges M. Bojowald for useful discussions, and the referees for recommending some improvements in the exposition.

\end{acknowledgement}


\end{document}